\documentclass[osajnl,twocolumn,showpacs,superscriptaddress,10pt]{revtex4-1} 
\usepackage{amsmath,amssymb,graphicx}
\begin{document}

\title{Simultaneous detection of rotational and translational motion in optical tweezers by measurement of backscattered intensity \\ 
}
\author{Basudev Roy}
\author{Sudipta K. Bera}
\author{Ayan Banerjee}
\affiliation{Department of Physical Sciences, IISER-Kolkata, Mohanpur 741252, India}
\email{Corresponding author: ayan@iiserkol.ac.in}

\begin{abstract}
We describe a simple yet powerful technique of simultaneously measuring both translational and rotational motion of mesoscopic particles in optical tweezers  by measuring the backscattered intensity on a quadrant photodiode (QPD). While the measurement of translational motion by taking the difference of the backscattered intensity incident on adjacent quadrants of a QPD is well-known, we demonstrate that rotational motion can be measured very precisely by taking the difference between the diagonal quadrants. The latter measurement eliminates the translational component entirely, and leads to a detection sensitivity of around 50 mdeg at S/N of 2 for angular motion of a driven micro-rod. The technique is also able to resolve the translational and rotational Brownian motion components of the micro-rod in an unperturbed trap, and can be very useful in measuring translation-rotation coupling of micro-objects induced by hydrodynamic interactions.
\end{abstract}

\maketitle 

Optical tweezers offer great versatility in the study of mesoscopic systems since they can exert controllable forces and torques on the trapped species. Particles trapped in optical tweezers execute both translational and rotational Brownian motion - however, the latter does not have any measurable signature for symmetric, non-birefringent particles. The rotational Brownian motion is more pronounced for particles having birefringence \cite{fri98}, or in asymmetric particles due to unbalanced scattering forces, which also may cause spontaneous rotation \cite{apl}. Precise detection of the rotational motion is therefore extremely important to determine the associated torque both for measurement as well as for application and control. In addition, hydrodynamic interactions in a viscous fluid  induce a coupling between the translational and rotational component of motion of a trapped micro-object \cite{martin} which can be measured to obtain very useful information about the nature of the interactions.  Now, translational motion is routinely detected by using fast cameras or by position sensitive detectors \cite{yodh}  which can quantify the change in the light scattered by a particle as it moves. Rotational motion can be differentiated into motion in a circular trajectory, and spinning about the center of mass of the particle. The former has components in the $x$ and $y$ directions which can be analyzed using cross-correlation techniques \cite{volpe} to determine the trajectory. On the other hand, spinning is more difficult to detect and, other than a dynamic light scattering based technique for anisotropic particles \cite{berne}, is mostly performed using polarization-based techniques presently. These include detecting the change in angular momentum of a circularly polarized trapping beam due to interaction with a rotating birefringent trapped particle \cite{dholakia}, or by detecting the intensity modulation in a polarization orthogonal to the input polarization due to excitation of plasmonic resonances in metal nano-particles \cite{orrit11}. Such techniques have also been extended to observe rotational Brownian motion in birefringent particles \cite{halina}, or in gold nanorods \cite{orrit11}.  However, such polarized-based techniques would thus have low sensitivity if the plasmonic or birefringence properties are weak. Also, the above-mentioned techniques usually do not address the simultaneous measurement of the translational and rotational motional components of motion. Such a measurement is indeed reported in Ref.~\cite{martin} with the use of a cross-correlation based technique, but it again relies on polarization properties of birefringent particles, and would thus not work for non-birefringent species.

Here we report an alternative approach that relies on differential detection of the back-scattered intensity from an asymmetric particle without relying on its birefringence or plasmonic properties. The backscattered intensity is measured on a quadrant photodiode (QPD) to extract both the translational and rotational component of motion simultaneously. While measuring the translation component of motion using QPD is very standard \cite{block04}, we demonstrate that the rotational component can also be measured very accurately while almost entirely suppressing the translational one. We perform experiments on a polymer micro-rod, and demonstrate sensitivity of around 50 mdeg in measuring angular motion in a driven optical trap, thus giving unprecedented S/N in detection of rotational motion. We also show that the technique is able to measure both translational and rotational Brownian motion of the micro-rod in a constant optical trap, and could thus have intensive applications in measurements of hydrodynamic interactions manifested in the beating of cilia in bacteria \cite{martin, friedrich}, or the motion of trapped objects in active systems \cite{cicuta}. 

\begin{figure}[!h!t]
{\includegraphics[scale=0.3]{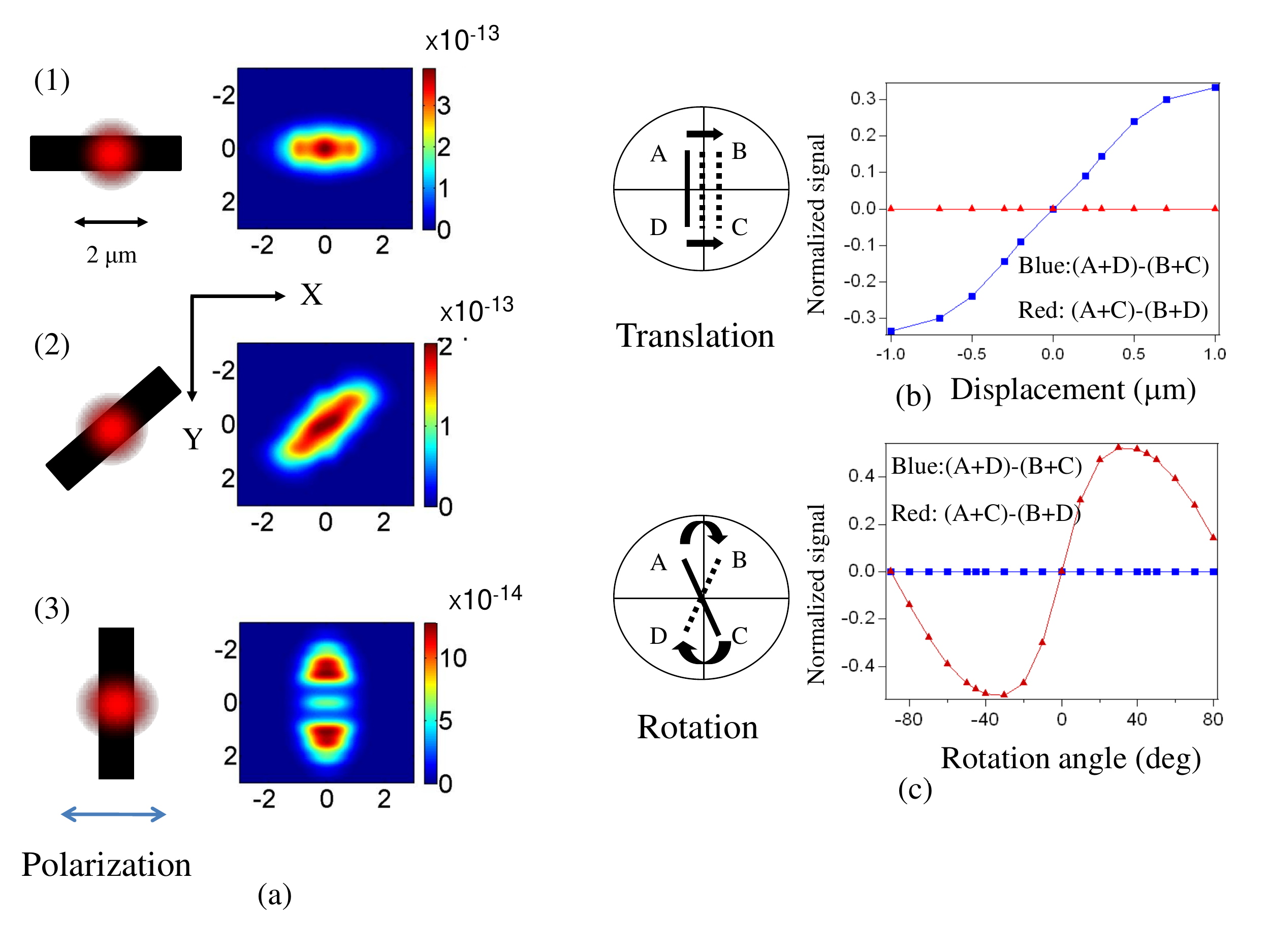}}
\caption[]{(Color online)(a) (1), (2), and (3) show a rod of dimension $4\times 1~ \mu$m, illuminated at the center by a laser beam of diameter $1~ \mu$m, oriented at (1) 0$\deg$, (2) 45$\deg$, and (3) 90$\deg$ to the polarization direction ($x$ axis) of the beam. Corresponding scatter patterns showing scattered field intensity are shown adjacent to the rod ($x$ and $y$ axes are in $\mu$m). (b) Net signal from a QPD as the rod is translated by 1 $\mu$m across the laser beam. The blue curve shows $S_{trans}$, while the red curve shows $S_{rot}$. (c) The net signal from a QPD as the rod is rotated by 90 $\deg$ across the laser beam.
\label{fig1}}
\end{figure}

The measurement of motion of particles in a QPD relies on changes in the scattered intensity of light from the particle due its motion with respect to a detection beam focused on it. For example, light scattered in the backward direction off a rod shaped particle shows an asymmetric intensity profile as shown in Fig.~\ref{fig1}(a) (rod dimensions $4\times 1~ \mu$m, scatter pattern calculated using Lumerical - a commercial FDTD-based software). The rod is aligned along the x axis and placed in a focused laser beam having a waist size of 1 $\mu$m. As the particle exhibits rotation and translation, the scattered pattern is modified likewise (Fig.~\ref{fig1}(a) 1 - 3). The scatter pattern can be detected using a QPD, which has four quadrants A-D, as shown in Fig.~\ref{fig1}. The translational motion of the rod in the x axis is obtained by the operation $S_{trans} = (A-B)+(D-C) = (A+D)-(B+C)$. On the other hand, if we perform $S_{rot} = (A-B)-(D-C) = (A+C)-(B+D)$, the common mode of the motion, i.e. the translational motion in this case, gets canceled. For rotational motion of the rod - which results in the rotation of the scatter pattern about the center of the QPD as well - the motion recorded in A-B is anti-correlated to the motion along D-C. Thus the rotational signal gets amplified upon performing $S_{rot}$, which is nothing but the difference of the sum of the diagonal quadrants.  As we show in Fig.\ref{fig1}(b), $S_{trans}$ changes from a positive to negative maxima for translational motion whereas $S_{rot}$ is zero, while $S_{rot}$ shows a cyclical behaviour for rotational motion with $S_{trans}$ being zero in Fig.~\ref{fig1}(c). Also, both $S_{trans}$ and $S_{rot}$ display a linear behaviour for small values of translation and rotation, respectively.

\begin{figure}[!h!t]
{\includegraphics[scale=0.2]{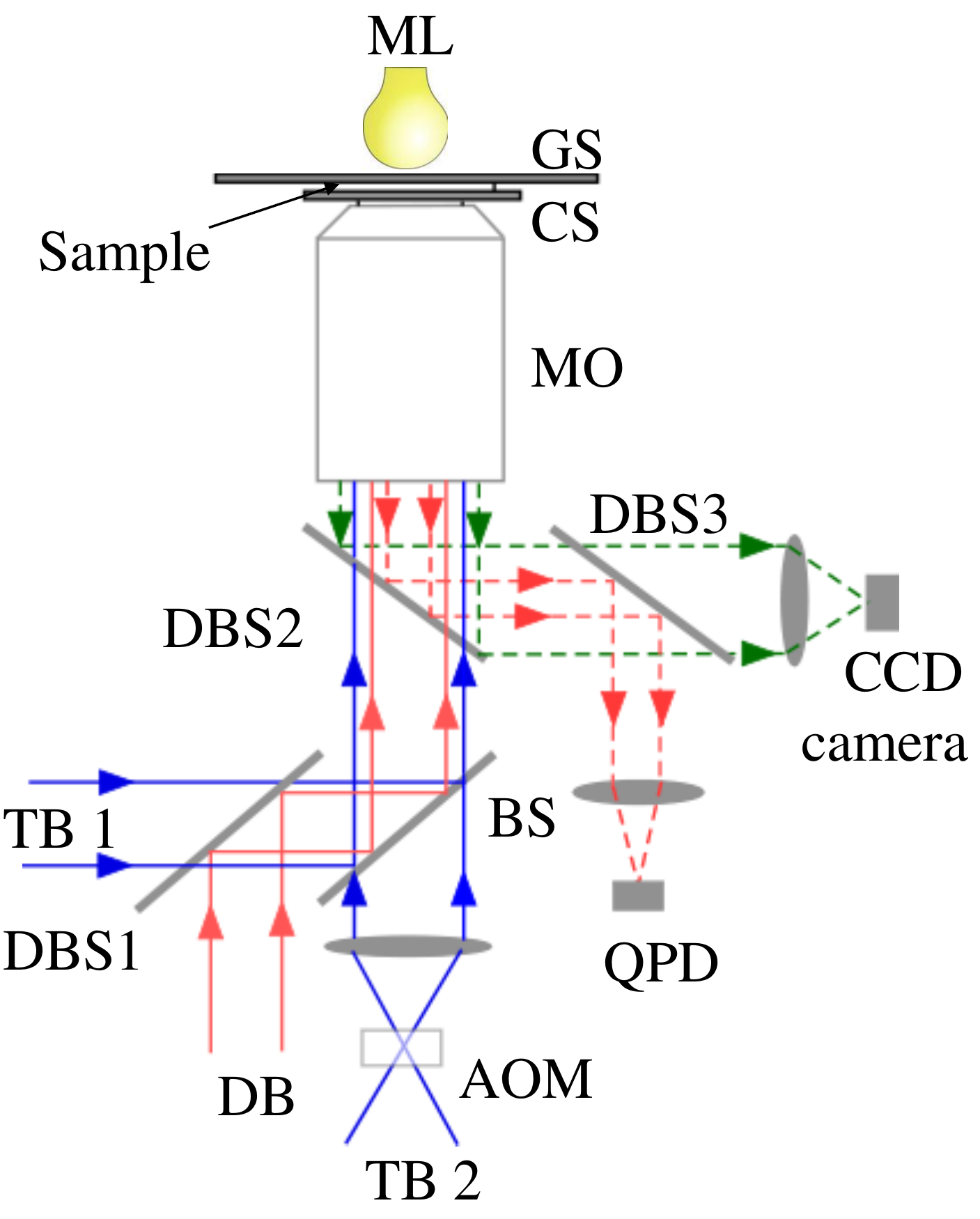}}
\caption[]{(Color online) Schematic of the experiment.  ML: Microscope lamp, GS: Glass slide, CS: Cover slip, MO: Microscope objective, DBS: Dichroic BeamSplitter, BS: 50-50 Beam splitter, TB: Trapping beam, DB: Detection beam, QPD: Quadrant photodiode, AOM: Acousto-Optic Modulator.
\label{fig2}}
\end{figure}

\begin{figure}[!h!t]
{\includegraphics[scale=0.24]{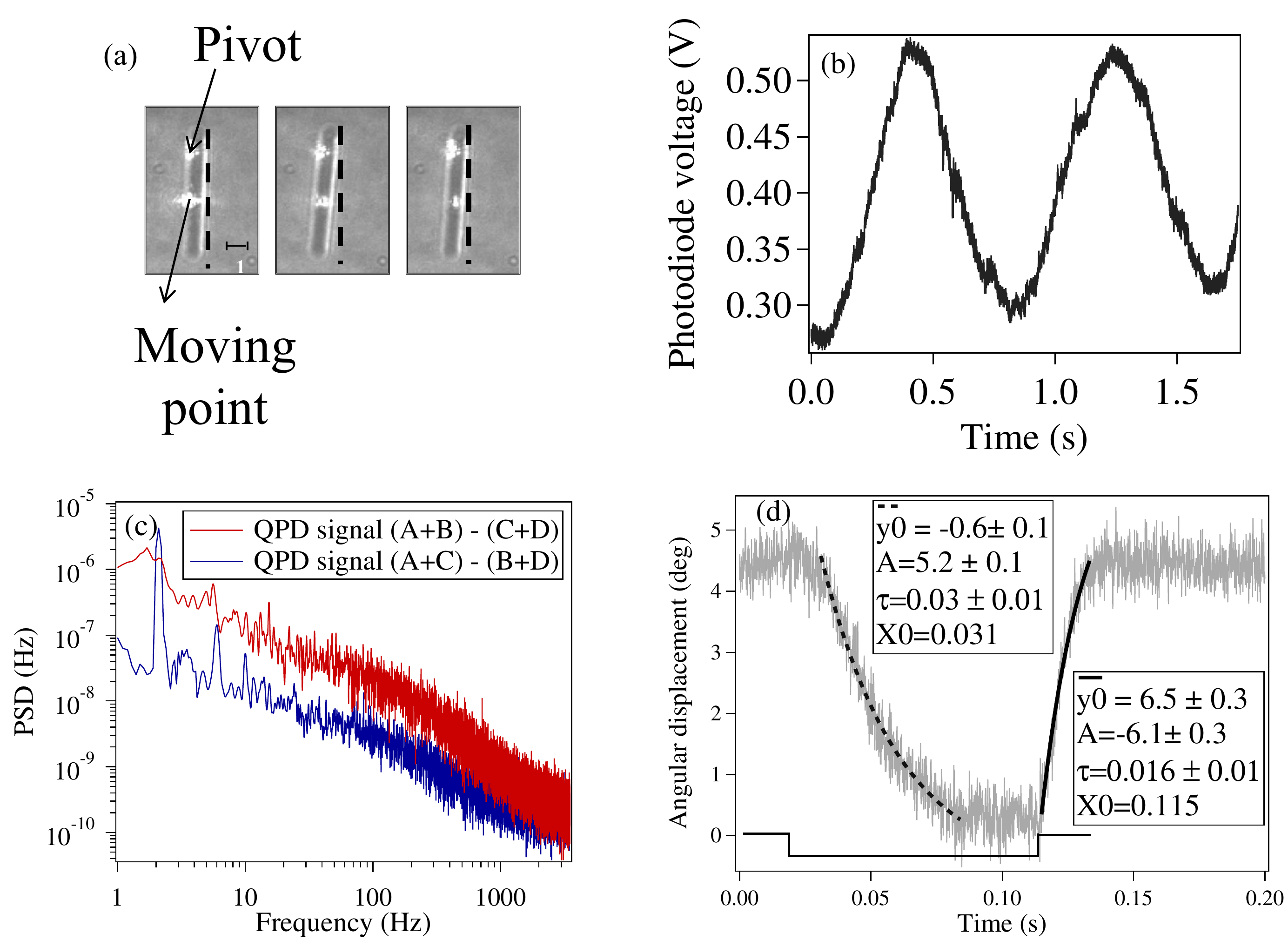}}
\caption[]{(Color online) (a) 1 - 3: Rotational motion of a rod ($4\times 1~\mu$m) using two traps - one near the top serving as pivot point, and the other lower down that was spatially modulated by the AOM (dimensions in microns). (b) QPD signal for $S_{rot}$. (c) Power spectra (PSD: Power spectral density) for $S_{trans}$ and $S_{rot}$. $S_{rot}$ shows peaks at the rod rotation frequency and higher harmonics generated due to the square wave. (d) Exponential response of the rod due to the square wave excitation (shown below the response). The time constants of the rising and falling exponentials are different due to the different trap stiffnesses at the two end locations. 
\label{fig3}}
\end{figure}

Our experimental system is shown in Fig.~\ref{fig2}. It is based on a standard optical tweezers built around a Carl Zeiss Axiovert.A1 inverted microscope with the trapping lasers at 1064 nm, and detection laser at 670 nm. A small volume (around 30 $\mu$l) of an aqueous dispersion of polymer micro-rods of size varying from $3\times 0.5$ to $6\times 1.5~ \mu$m was taken in a sample chamber made out of a cover slip stuck by double-sided tape to a microscope glass slide. In the experiment, a polymer micro-rod was held at two points by two trapping beams focused using an objective lens of numerical aperture (NA) 1.4. One of the beams was taken from the first order of an acousto-optic modulator (AOM) and was moved orthogonal to the axis of the rod, while the other beam was kept fixed, thereby acting as a pivot.  This configuration gave the rod a small rotational motion. Figure \ref{fig3}(a) shows a series of 3 images of such rotational motion of a trapped rod of dimension $4\times 1~\mu$m as the beam from the AOM was moved. The rotation amplitude was around 5 degrees, which we calibrate from a frame by frame analysis of the rod motion, using a 1 $\mu$m polystyrene bead as reference. The detection laser beam was overlapped with the pivot beam and the change in scatter profile due to the motion of the particle was imaged on a QPD. The QPD we use has been extracted from a CD player head (chip Sony KSS-213C) (see Ref.~\cite{sam12} for details) and gives independent voltage outputs for each of the 4 quadrants. The QPD signals are recorded using a National Instruments PCIe-6361 data card, and the data was averaged for 5s at a sampling rate of 40 kHz.

The first order beam from the AOM was moved periodically by modulating the voltage controlled oscillator (VCO) voltage and both $S_{trans}$ and $S_{rot}$ were measured simultaneously. A typical trace of $S_{rot}$ for sinusoidal modulation is shown in Fig.~\ref{fig3}(b). The quantitative differences in $S_{trans}$ and $S_{rot}$ are understood by determining the power spectra of the signals as we show in Fig.~\ref{fig3}(c), in which case a square wave of frequency 2 Hz was used to modulate the VCO. We find that the noise floor drops down by about one order of magnitude in $S_{rot}$ (blue curve), while peaks appear at the rotation frequency of 2 Hz, as well as higher odd harmonics since we modulate with a square wave, which demonstrates the sensitivity of this technique. The S/N of the fundamental rotation peak was estimated by taking the  ratio of height of the peak in the power spectrum and the average of the noise floor in the vicinity of the peak, and was calculated to be around 200. Since this corresponded to a rotation amplitude of 5 degrees, the detection sensitivity for this micro-rod could be quantified to 50 mdeg at a S/N of 2.  
When the AOM frequency is moved as a square wave, the equilibrium orientation shifts suddenly and the rod follows. The dynamics is given by the following equation 
\begin{equation}
 D_{t} \frac{d \theta}{d t} + k_1 (\theta - \theta_0) = C \eta (t)
\end{equation}
Here, the $D_{r}$ is the rotational drag coefficient, k the trap stiffness and $\eta(t)$ the delta correlated noise term that generates random motion in the angular coordinate, and $C$ is a constant. This is satisfied by a decaying exponential from one equilibrium position to another. Thus, as the rod gets closer to the equilibrium orientation, its velocity becomes slower. The time constant of this process is $\dfrac{-D_{r}}{k_1}$ and is inversely proportional to the intensity of the laser at the trapping spot. Figure \ref{fig3}(d) shows the exponential decay of the angular orientation of the rod given by $S_{rot}$ as the equilibrium point is shifted suddenly. We fit to a function $y = y_0 + A\exp {(x-x_0)/\tau}$, where we ignore the effects of the stochastic noise term of the Langevin equation since each trace is the average of 20 individual datasets collected for 5s so that the effects of the random fluctuations are reduced considerably. Even otherwise, the fluctuations would merely have introduced a noise band around the exponential without changing its nature. We find that there are two different time constants in the signal while going from one equilibrium position to the other. This results from the difference in the values of the trapping power at the extreme ends of the moving beam due to the reduced efficiency of the first order of the AOM as the VCO voltage is modified. Thus, the trapping power at the start position of the rod is 15 mW, which falls to 8 mW at the end position. Now, since the time constant depends inversely upon the angular trap stiffness, $\tau$ for the decaying exponential is $0.03\pm 0.01$ s, while that for the rising one is $0.016 \pm 0.01$ s, which agrees well with the ratio of trapping stiffnesses at the two ends, and gives a good consistency check to our measurement technique. Using the value of the drag coefficient for a cylindrical rod of the dimensions shown in Fig.~\ref{fig3}(a), we find that the angular trap stiffness is 10 nN nm/rad.  

\begin{figure}[!h!t]
{\includegraphics[scale=0.2]{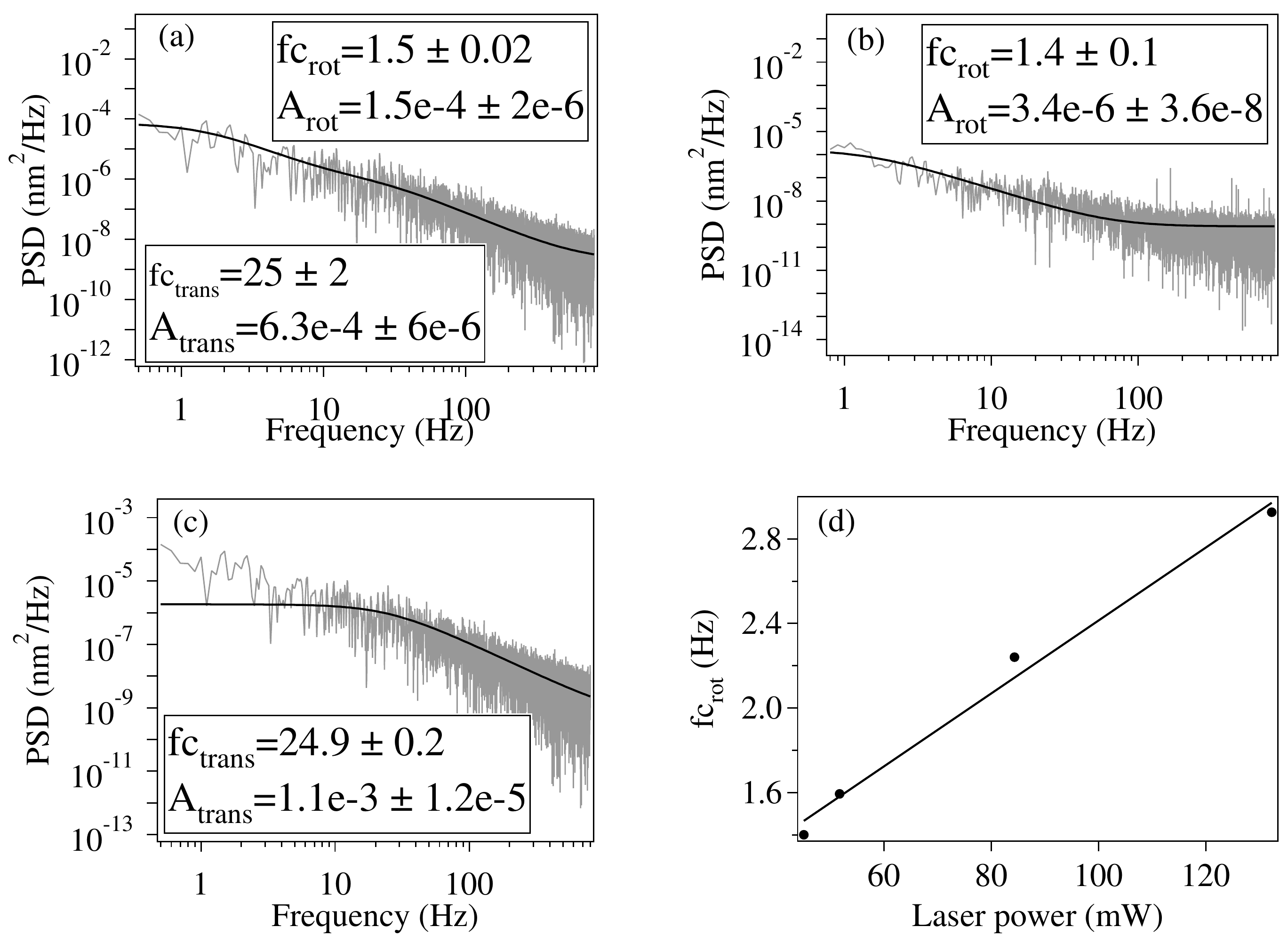}}
\caption[]{(Color online) (a) Power spectrum of $S_{trans}$ with double Lorentzian fit. (b) Power spectrum of $S_{rot}$ with Lorentzian fit. (c) Power spectrum of $S_{trans}$ with Lorentzian fit from $x=5$ Hz so as to completely eliminate the rotational component whose corner frequency is 1.5 Hz. (d) Variation of $fc_{rot}$ with laser beam power (intensity).
\label{fig4}}
\end{figure}

Note that any misalignment of the detection beam with the pivot point, or any translation of the pivot point itself, leads to coupling of rotational motion with the translational one, which implies that there would be a greater response obtained at the modulation frequency in the power spectrum of the $S_{trans}$ signal.  The power spectrum of $S_{trans}$ thus yields a double Lorentzian that can be understood as the solution to a set of coupled Langevin equations, one for translational and the other for rotational Brownian motion. A typical set of equations are given as follows  
\begin{equation}
 d x + k_1x dt + k_2\theta dt = \sqrt{2 D_t dt} \eta_1
\end{equation}
\begin{equation}
 d \theta + k_3 \theta dt + k_4 x dt = \sqrt{2 D_r dt} \eta_2
\end{equation}

where, $k_1$ and $k_3$ are the translational and rotational trap stiffnesses and, $k_2$ and $k_4$ are the cross-coupling coefficients. The $D_t$ and $D_r$ are the corresponding diffusion coefficients. Solving this set of coupled differential equations yields the time series for the translational  and rotational positions of the particle. Since the rotational and translational processes appear in the system together with a weak coupling between them, they give rise to independent Lorentzian responses simultaneously which leads to a double Lorentzian in the power spectra, similar to that described in Refs.~\cite{volpe2, curran} (data not shown). Figure \ref{fig4}(a) shows such a double Lorentzian obtained experimentally giving a rotational corner frequency $fc_{rot}$ of $1.5\pm 0.01$ Hz and a translational corner frequency $fc_{trans}$ of $15\pm 1.2$ Hz. To verify the efficacy of our technique, we compare these values to the corner frequencies obtained by fitting to the power spectra for the individual components separately. For the rotational component, we fit to the power spectrum of $S_{rot}$ (Fig.~\ref{fig4}(b)), which shows an order of magnitude reduction in amplitude compared to $S_{trans}$ similar to Fig.~\ref{fig3}(c). The corner frequency obtained is $1.4\pm 0.1$ Hz, which is in very close agreement to that from Fig.~\ref{fig4}(a). It is also clear that the translational component is almost entirely eliminated in this measurement, which proves its usefulness in measuring rotational motion. For measuring $fc_{trans}$ separately, we revert back to the power spectrum of $S_{trans}$, but start the fitting from 5 Hz so as not to pick up the rotational component at all. The value of $fc_{trans}$ we thus obtain is $14.8\pm 0.2$ Hz, which once again agrees with the value obtained from the double Lorentzian fit. The fit amplitudes $A_{trans}$ (which give the values for the diffusion constants after normalizing with the QPD sensitivity \cite{block04}) also agree well for the translational power spectra for double and single Lorentzian fits.  However, for the rotational case, the $A_{rot}$ values from the double lorentizan and that from power spectrum of $S_{rot}$ are very different since the QPD sensitivities for the signal modes vary substantially. In addition, we also measure $fc_{rot}$ as a function of laser power, which yields a straight line as is shown in Fig.~\ref{fig4}(d). This is quite expected since just as the translational Brownian motion corner frequency increases linearly with intensity, the same can be expected for the rotational component \cite{halina}.

In conclusion, we have demonstrated simultaneous measurement of translational and rotational motion of an asymmetric particle (a polymer micro-rod) in optical tweezers by the use of a single quadrant photo diode without any change in experimental configuration. The voltage output of all four quadrants is available to us, and we take the difference of the sum of the adjacent quadrants for the translational motion, and that of the diagonal quadrants for the rotational motion. The latter is demonstrated for the first time, and is remarkably sensitive in picking up rotational motion while eliminating the translational component. We demonstrate a sensitivity of around 50 mdeg for rotational motion of the micro-rod. Note that the sensitivity could be higher with particles having higher refractive index contrast with respect to the fluidic environment of the trap that would lead to enhanced scattering. While high sensitive detection of translational motion of trapped particles using a QPD has already been demonstrated \cite{perkinsopex}, our design could facilitate the measurement of rotational motion with the same levels of sensitivity without relying on polarization properties of the scattering particle. Such simultaneous detection of rotational and translational motion could be of use in the measurement of hydrodynamic interactions where cross-correlation studies could lead to greater understanding of the complex motion of cilia and flagella. 

This work was supported by the Indian Institute of Science Education and Research, Kolkata, an autonomous research and teaching institute funded by the Ministry of Human Resource Development, Govt. of India. SKB  acknowledges the Council of Scientific and Industrial Research, Govt. of India for a research fellowship.

\vspace*{5mm}
\newpage
{\bf Detailed Reference list:}
\begin{enumerate}
\item M. E. J. Friese, T. A. Nieminen, N. R. Heckenberg and H. Rubinsztein-Dunlop, {\it Nature}, ``Optical alignment and spinning of laser-trapped microscopic particles", {\bf 394}, 348 (1998).
\item P. Galajda and P. Ormos, ``Complex micromachines produced and driven by light", {\it Appl. Phys. Lett.} {\bf 78}, 249 (2001).
\item S. Martin, M. Reichert, H. Stark and T. Gisler, ``Direct observation of hydrodynamic rotation-translation coupling between two colloidal spheres", {\it Phys. Rev. Lett.} {\bf 97}, 248301 (2006).
\item Y. Han, A. M. Alsayed, M. Nobili, J. Zhang, T. C. Lubensky and A. G. Yodh, ``Brownian motion of an ellipsoid", {\it Science} {\bf 314}, 626 (2006).
\item G. Volpe and D. Petrov, ``Torque detection using Brownian fluctuations", {\it Phys. Rev. Lett.} {\bf 97}, 210603 (2006).
\item  B. J. Berne and R. Pecora, {\it Dynamic Light Scattering}, (Dover, New York), (2000).
\item Y. Arita, M. Mazilu and K. Dholakia, ``Laser induced rotation and cooling of a trapped microgyroscope in vacuum", {\it Nature Comm.} {\bf 4}, 2374 (2013).
\item P. V. Ruijgrok, N. R. Verhart, P. Zijlstra, A. L. Tchebotareva and M. Orrit, ``Brownian fluctuation and heating of an optically trapped gold nanorod", {\it Phys. Rev. Lett.} {\bf 107}, 037401 (2011).
\item J. S. Benett, L. J. Gibson, R. M. Kelly, E. Brousse, B. Baudisch, D. Preece, T. A. Nieminen, T. Nicholson, N. R. Heckenberg, H. Rubinsztein-Dunlop, ``Spatially-resolved rotational microrheology with an optically trapped sphere", {\it Sci. Rep.} {\bf 3}, 1759 (2013).
\item K. C. Neuman and S. M. Block, ``Optical trapping", {\it Rev. Sci. Instrum.} {\bf 75}, 2787 (2004).
\item V. F. Geyer, F. Julicher, J. Howard and B. F. Friedrich, ``Cell-body rocking is a dominant mechanism for flagellar synchronization in a swimming alga", {\it Proc. Natl. Acad. Sci. U.S.A.} {\bf 110}, 18058 (2013).
\item J. Kotar, M. Leoni, B. Bassetti, M. C. Lagomarsino and P. Cicuta, ``Hydrodynamic synchronization of colloidal oscillators", {\it Proc. Natl. Acad. Sci. U.S.A.} {\bf 107}, 7669 (2010).
\item S. B. Pal, A. Haldar, B. Roy and A. Banerjee, ``Measurement of probe displacement to the thermal resolution limit in photonic force microscopy using a miniature quadrant photodetector", {\it Rev. Sci. Instrum} {\bf 83}, 023108 (2012).
\item G. Volpe and D. Petrov, "Torque detection using Brownian fluctuations",  {\it Phys. Rev. Lett.}, {\bf 97}, 210603 (2006).
\item  A. Curran, M. P. Lee, M. J. Padgett, J. M. Cooper and R. Di Leonardo, ``Partial synchronization of stochastic oscillators through hydrodynamic coupling" {\it Phys. Rev. Lett.} {\bf 108}, 240601 (2012). 
\item A. R. Carter, G. M. King, and T. T. Perkins, ``Back-scattered detection provides atomic-scale localization precision, stability, and registration in 3D", {\it Opt. Exp.} {\bf 15}, 13434 (2007).

\end{enumerate}


\begin{thebibliography}{9}
{\small
\bibitem{fri98} M. E. J. Friese, T. A. Nieminen, N. R. Heckenberg and H. Rubinsztein-Dunlop,
 {\it Nature}, {\bf 394}, 348 (1998).

\bibitem{apl} P. Galajda and P. Ormos {\it Appl. Phys. Lett.} {\bf 78}, 249 (2001).

\bibitem{martin} S. Martin, M. Reichert, H. Stark and T. Gisler, {\it Phys. Rev. Lett.} {\bf 97}, 248301 (2006).

\bibitem{yodh} Y. Han, A. M. Alsayed, M. Nobili, J. Zhang, T. C. Lubensky and A. G. Yodh, {\it Science} {\bf 314}, 626 (2006).

\bibitem{volpe} G. Volpe and D. Petrov, {\it Phys. Rev. Lett.} {\bf 97}, 210603 (2006).

\bibitem{berne} B. J. Berne and R. Pecora, {\it Dynamic Light Scattering}, (Dover, New York), (2000).

\bibitem{dholakia} Y. Arita, M. Mazilu and K. Dholakia, {\it Nature Comm.} {\bf 4}, 2374 (2013).

\bibitem{orrit11} P. V. Ruijgrok, N. R. Verhart, P. Zijlstra, A. L. Tchebotareva and M. Orrit, {\it Phys. Rev. Lett.} {\bf 107}, 037401 (2011).

\bibitem{halina} J. S. Benett, L. J. Gibson, R. M. Kelly, E. Brousse, B. Baudisch, D. Preece, T. A. Nieminen, T. Nicholson, N. R. Heckenberg, H. Rubinsztein-Dunlop, {\it Sci. Rep.} {\bf 3}, 1759 (2013).

\bibitem{block04} K. C. Neuman and S. M. Block, {\it Rev. Sci. Instrum.} {\bf 75}, 2787 (2004).

\bibitem{friedrich} V. F. Geyer, F. Julicher, J. Howard and B. F. Friedrich, {\it Proc. Natl. Acad. Sci. U.S.A.} {\bf 110}, 18058 (2013).

\bibitem{cicuta} J. Kotar, M. Leoni, B. Bassetti, M. C. Lagomarsino and P. Cicuta, {\it Proc. Natl. Acad. Sci. U.S.A.} {\bf 107}, 7669 (2010).

\bibitem{sam12} S. B. Pal, A. Haldar, B. Roy and A. Banerjee, {\it Rev. Sci. Instrum.} {\bf 83}, 023108 (2012).

\bibitem{volpe2} G. Volpe and D. Petrov, {\it Phys. Rev. Lett.}, {\bf 97}, 210603 (2006).

\bibitem{curran} A. Curran, M. P. Lee, M. J. Padgett, J. M. Cooper and R. Di Leonardo, {\it Phys. Rev. Lett.} {\bf 108}, 240601 (2012). 

\bibitem{perkinsopex} A. R. Carter, G. M. King, and T. T. Perkins, {\it Opt. Exp.} {\bf 15}, 13434 (2007).
}
\end{thebibliography}
\end{document}